# An extended watershed-based zonal statistical AHP model for flood risk estimation: Constraining runoff converging related indicators by sub-watersheds

**Hongping Zhang[a], Zhenfeng Shao[a, *], Jinqi Zhao[a,b], Xiao Huang[c], Jie Yang[a], Bin Hu[a], Wenfu Wu[a]**

[a] State Key Laboratory for Information Engineering in Surveying, Mapping and Remote Sensing, Wuhan University, Wuhan 430079, China;

yx_zhping@126.com (H.Z.); shaozhenfeng@whu.edu.cn (Z.S.); masurq@whu.edu.cn (J. Z); yangj@whu.edu.cn (J.Y.); hubin763259288@whu.edu.cn (B.H.); wuwwf09140818@whu.edu.cn (W.W.).

[b] Jiangsu Key Laboratory of Resources and Environmental Information Engineering, China University of Mining and Technology, Xuzhou 221116, China.

[c] Department of Geosciences, University of Arkansas, Fayetteville, AR 72701, USA; xh010@uark.edu

[*] Corresponding author: shaozhenfeng@whu.edu.cn (Z.S.)

**Abstract:** Floods are highly uncertain events, occurring in different regions, with varying prerequisites and intensities. A highly reliable flood disaster risk map can help reduce the impact of floods for flood management, disaster decreasing, and urbanization resilience. In flood risk estimation, the widely used analytic hierarchy process (AHP) usually adopts pixel as a basic unit, it cannot capture the similar threaten caused by neighborhood source flooding cells at sub-watershed scale. Thus, an extended watershed-based zonal statistical AHP model constraining runoff converging related indicators by sub-watersheds (WZSAHP-Slope & Stream) is proposed to fill this gap. Taking the Chaohu basin as test case, we validated the proposed method with a real-flood area extracted in July 2020. The results indicate that the WZSAHP-Slope & Stream model using multiple flow direction division watersheds to calculate statistics of *distance from stream* and *slope* by maximum statistic method outperformed other tested methods. Compering with pixel-based AHP method, the proposed method can improve the correct ratio by 16% (from 67% to 83%) and fit ratio by 1% (from 13% to 14%) as in validation 1, and improve the correct ratio by 37% (from 23% to 60%) and fit ratio by 6% (from 12% to 18%) as in validation 2.

## *1. Introduction*

Flooding is a common phenomenon occurring worldwide, related to climatic conditions, geography, the environment, human activities, and other factors. Flooding is a kind of natural event with great destructive power and a common challenge for human society (Teng et al., 2017; Wheater, 2006). The underlying surface is a formation environment of floods, and it is also the main sites for human activity. As economic development and urbanization promotion, more and more natural pervious surface is replaced by impervious surface. The increase in impervious surfaces may not only cause an increase in surface runoff, but also to a decrease in the time between the start of a flood and the flood peak (Shao et al., 2019). These negative hydrological effects may cause an increase in extreme rainstorm events in the future. Continuous extreme rainfall will cause the water level of rivers, lakes and reservoirs to rise sharply, this will bring out serious threaten to flood prone areas (Bertilsson et al., 2019). This is especially true in China.

China is one of the countries frequently disturbed by flood and water logging events. It was reported that there are 641 out of 654 of Chinese cities exposed to frequent floods (Jiang et al., 2018). Taking Chinese third largest catchment--the Yangtze River as an example, the areas located in the middle and lower reaches of this estuary experience plum floods every spring. In 2020, there were five flooding whose discharge or water level reached the flood warning level for the Yangtze river. In the flood season of July, 2020, there were seven floods taking place across several provinces. According to the Ministry of Emergency Management of China, as of 13 August 2020, the floods had affected 63.46 million people and caused a direct eco- nomic loss of 178.96 billion CNY (Wang et al., 2021). Different types of human settles might experience different level of loss in the same hazard (Xu et al., 2020). The combination of climate change and increasing urbanization, brings great challenges to planning and managing cities for sustainability (Arduino et al., 2005, Bertilsson et al., 2019). Meanwhile, the feedback loop of flood protection and the flood awareness of community affect their planning and development in long term (Barendrecht et al., 2017). In order to mitigate urban floods, accurate flood risk estimation corresponding to patterns of urbanization is a policy objective.

Flood risk estimation aims to identify active areas as extreme discharges expose communities to dangerous situations. Flood risk estimation focused on three factors: the flood risk distribution, extreme weather and climate events, and risk exposure in a flood affected area. Among these, the flood risk distribution is the dominant factor. The climate conditions and socio-economic factors are external influences, interacting with each other and altering flood risks. Flood risk maps support identification and mitigation of those risks.

Numerical simulation and the decision-making methods are widely used when mapping flood risk areas. Numerical simulation methods such as LISFLOOD (Bates and De Roo, 2000; Coulthard et al., 2013; De Roo et al., 2000; Saynor et al., 2018) and HEC-RAS (Pasquier et al., 2018) rely on detailed input data such as rainfall, evaporation, sunshine time, topography, soil, river, and drainages to estimate flood risk, but are computationally demanding. Given this limitation, qualitative flood risk evaluations via decision-making models such as analytic hierarchy process (AHP) (Haider et al., 2019), Bayesian models (Han and Coulibaly, 2017), fuzzy comprehensive evaluation (Tang et al., 2018), swarm intelligence and evolutionary algorithm (Lai et al., 2016) have been deployed. Most of these methods depend on situ flood records, and may under-estimate the likelihood more serious floods in the future. Bayesian models need measurements and statistical data from historical flood events. Fuzzy methods and swarm intelligence can be considered as supervised classification, but need samples from historical flood events. Flood risk estimation via AHP method does not rely directly on historical records.

The flood risk estimation via AHP models constructs multi-level criteria with a series of reasonable weights to reflect individual target features, and therefore are widely used to generate flood risk maps. Abdouli et al. (2019) adopted AHP to map the flooding potential on the Arabian Gulf coast using data that include land use, soil type and antecedent moisture. Wu et al. (2019) estimated flood vulnerability via AHP using rainfall intensity and duration, elevation, slope, land use, population density. In flood risk estimation via AHP models, it is essential to describe the individual hydrological heterogeneity features, as well as approximate risk characteristics of neighborhood geographic cells (Chakraborty and Mukhopadhyay, 2019; Ouma and Tateishi, 2014). However, these pixel-based AHP fail to consider terrain connectivity, thus ignoring consistency in flood-risk across neighborhood pixels.

We argue that using watershed as a basic unit will be more suitable and closer to the natural flood processes than traditional pixel-based AHP models. A watershed is a natural edge of the water that converges through the terrain. During a flooding event, the continuous rainwater converged along the flow path can be considered as a source flooding to its sub-watersheds. The flood risk of cells is more dependent on the maximum risk level of neighborhood cells in sub-watershed scale rather than the terrain features or hydrological characteristics of individual cells.

Therefore, we propose an extended watershed-based zonal statistical AHP model constraining runoff converging related indicators by sub-watersheds (WZSAHP-Slope & Stream). Taking Chaohu basin of Anhui province of China as an example, we validated the performance of our proposed model with real-world flood events occurred in July 2020, and compared the performance of the proposed method to the original AHP methods. Meanwhile, we discuss the influence of watershed divided by the classical single flow direction—the D8 algorithm and the multiple flow directions (MFD) algorithm which follows our previous work (Zhang et al., 2020).

We also discuss the performance of methods using every indicator and our proposed method using runoff related indicators and zonal statistics via sub-watersheds divided by MFD.

This paper is organized as follows: The Chaohu basin study area and the main data sources will be introduced in section 2. The methodology including the proposed model and the overall workflow of our study will be described in section 3. Subsequently, the results and the major findings, as well as the influence factors will be discussed in section 4. Some conclusions and the limitation of this study are provided in section 5

## *2. Study area and materials*

### *2.1. Study area*

The Chaohu basin is located at the intersection of the Yangtze River catchment and Huai River catchment of China, draining rainwater from the upstream Hua River. We selected counties that surround Chaohu Lake using a 10-kilometer buffer as the study area, as shown in Fig.1.

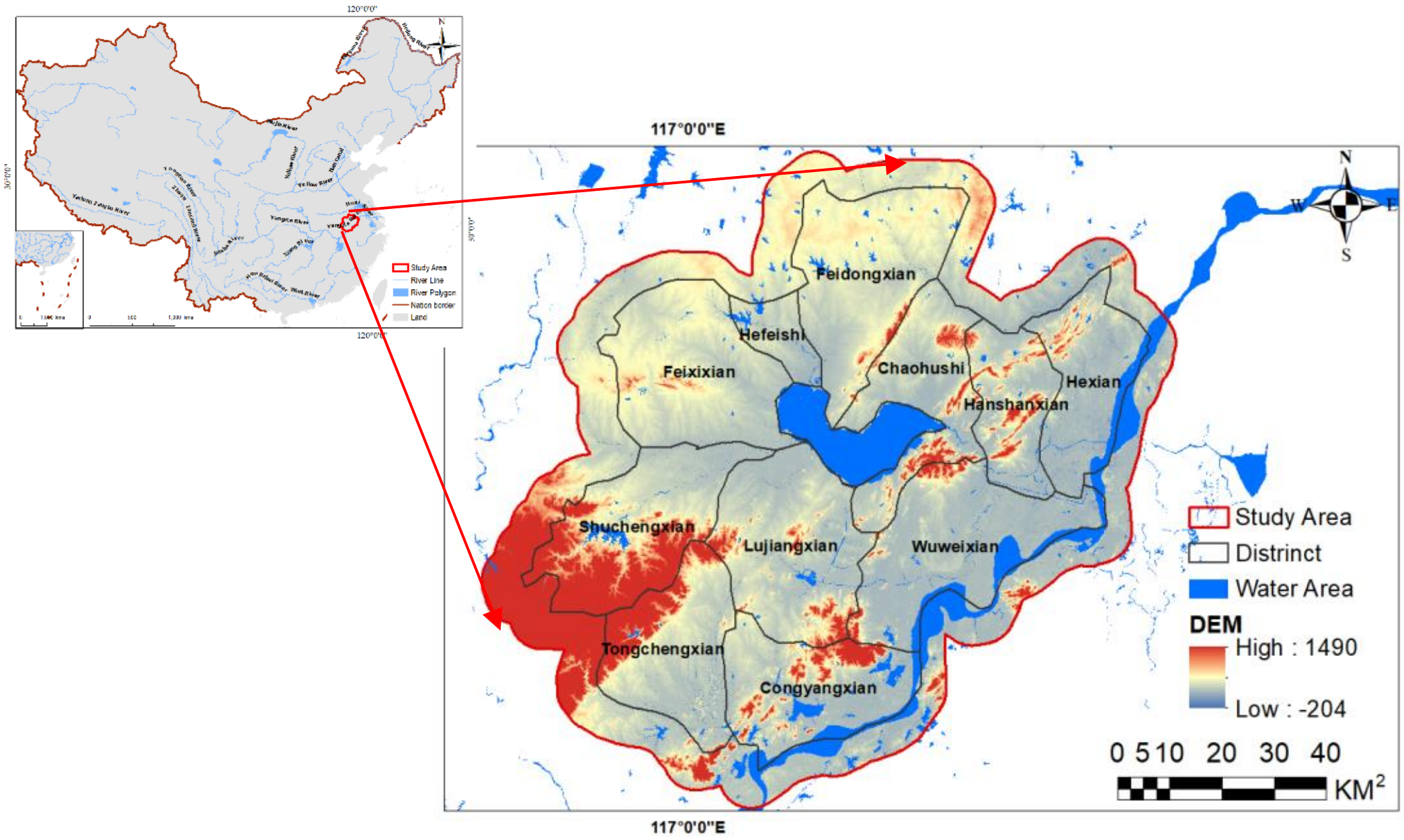

Figure 1. The geographical location of the study area and the DEM of the study area (according to the China basic geographic information, 2008 version).

The study area is characterized by a "butterfly" terrain pattern, containing five major terrain types: low mountains, hilly land, hill land, plains (lakeside and wavy plains) and water areas. The terrain in the northwest and southeast is higher than in other areas of the sudy area; the middle region has a relatively low elevation. Lujiang and Wuwei are at relatively lower elevations, and therefore, are exposed to a high risk of flood. During the Meiyu (June to August) period in 2020, the Chaohu basin reached the highest water level in history (13.48m) on July 22, 2020 due to continuous and intense rainfall. The record-breaking water levels remained for several days, posing a direct threat to the cities of Chaohu, Hefei. The local government decided to break the several dams, e.g., Chuhe, Quanjiao etc., to drain rainwater. A total of 188 (out of 288) low-lying village parcels along flow-in rivers in the Chaohu basin such as Fengle River, Feixi River, and the Zao River flooded at that time. Improving the accuracy of flood risk estimation in the Chaohu basin will support future flood risk management and thus help protect the local economy.

### *2.2. Materials*

Major data sources and their detailed information are presented in Table 1. We used a geographic information system (GIS) vector to map distinct and hydrological information. A digital elevation model (DEM) dataset was used to divide watersheds and calculate slopes. Impervious surface products were used to extract land cover and hydrological infiltration information. SAR (synthetic aperture radar) images were used to extract flooded areas. The pre-processing involving indicators flood risk estimation will be introduced as follows. The pre-processing steps for these materials involved in flood risk estimation included transforming, projecting, mosaicking and clipping, implemented in ArcGIS 10.3. We also used flood-related Baidu News to validate flooding that occurred at the town scale.

Table 1. Datasets used in this study.

| Data Sources | Used Data | Detailed Information |
|---|---|---|
| Geographic information (1:1million) | District | The county and town level districts were used. |
| | River and lake | Hydrological layers were used to constraint DEM. |
| ASTER GDEM V2 (30m) | DEM | The DEM is used to divide watersheds and classify the slope and elevation indicators. The DEM was downloaded from [*http://www.gscloud.cn*](http://www.gscloud.cn). |
| | Land use type | |
| China's impervious surface product (2m) | Hydrological characteristics | Following (Zhang et al., 2020), the water, vegetation, soil, building and road layers were used to classify land use and hydrological indicators. |
| Images for extracting flooding areas | Water bodies | We used Landsat 8 OLI on 2020-7-20 and GF-3 on 2020-7-24 to extract flooding areas. The Landsat 8 OLI was downloaded from *https://www.usgs.gov*. The GF-3 data is supported by the GaoFen center of Hubei province. |
| Flooding information in Baidu News | Flooding and dam breaks by towns | Baidu news, as a validation source, was searched from *http://news.baidu.com*. |

- **GIS Vector maps—Geographic information**. We used the vectorized county and town boundaries released in 2008. The administrative districts served as geographic constraints to filter Baidu News flood reports. The hydrological layers including rivers, streams, and lakes (levels 1 to 5) were compared with water bodies as classified from images.

- **DEM—ASTER GDEM V2 dataset**. The DEM dataset we adopted is ASTER GDEM V2. Our study area covers a total of nine scenes. The horizon resolution of the DEM dataset is 30m, and the vertical resolution is 1m. The projection is WGS_1984_UTM_Zone_50N.

- **GRID of Remote sensing production—Land information.** We adopted China's impervious surface grid product (2m) (Shao et al., 2019). The 18 tiles for Hefei, Luan, Anqing, Wuhu, Maanshan, Chuzhou, and Huainan cities were used to prepare the land use and hydrological indicators.

- **RS—Flooded area extraction.** The flooded areas were used to validate our estimated results. We merged the extracted water bodies from the Landsat 8 OLI image collected on July 20, 2020 and from the GF-3 image collected on July 24, 2020 as potential flood areas. The pre-processing of Landsat 8 OLI performed in ENVI 5.3, including radiometric calibration, atmosphere correction. We classified water bodies by maximum likelihood method. The pre-processing of GF-3 SAR images were implemented in PolSAR Pro 5.2 and GAMMA including speckle filtering, radiometric calibration, terrain correction and geo-coding. The water bodies in GF-3 were classified basing the threshold segmentation method corresponding to the water bodies in Landsat 8 OLI. We excluded water bodies extracted from China's impervious surface product. Considering the temporal differences between these two images, we conducted validation experiments using the intersecting area to establish the ground truth range.

- **Baidu News—Flood events (flood and dike rupture information)**. Damage information from floods and dikes of flood events in July 2020 were used to verify the accuracy of the flood risk estimation in the Chaohu basin. We used an internet context searching and capturing tool named "Octopus" *(https://www.bazhuayu.com)* to collect information from the Baidu News website *(https://news.baidu.com)*. The keywords were used to filter Baidu news including "flood" and "waterlogging" combined with the county names.

## *3. Methodology*

In flood risk estimation, using sub-watersheds as basic unit can consider the hydrological influence introduced by adjacent pixels. Watershed is a natural edge directly derived by the runoff converges path through the terrain. Sub-watershed can reflect inner cells collecting and converging rainwater into the same connected hydrological water system. Different division of watersheds significant affect the simulation results of rainwater converging discharge and the soil erosion deposition (Hembram and Saha, 2020; Hu et al., 2017). Corresponding to natural floods, these areas nearby the flood source may face similar flooded threaten (Shao et al., 2019). Flood risk introduced by neighborhood cells can be captured in same sub-watershed scale. However, as in flood risk estimation, AHP always adopts pixel as basic unit, hence this may fail to express similar flooded threaten brought out by source cells in sub-watershed scale. Therefore, to end this gap, we proposed a extend AHP model, termed WZSAHP-RC using sub-watershed as basic unit to calculate statistics value for original indicators that influence runoff-covering process. The technical workflow used in this study is illustrated in Fig.2.

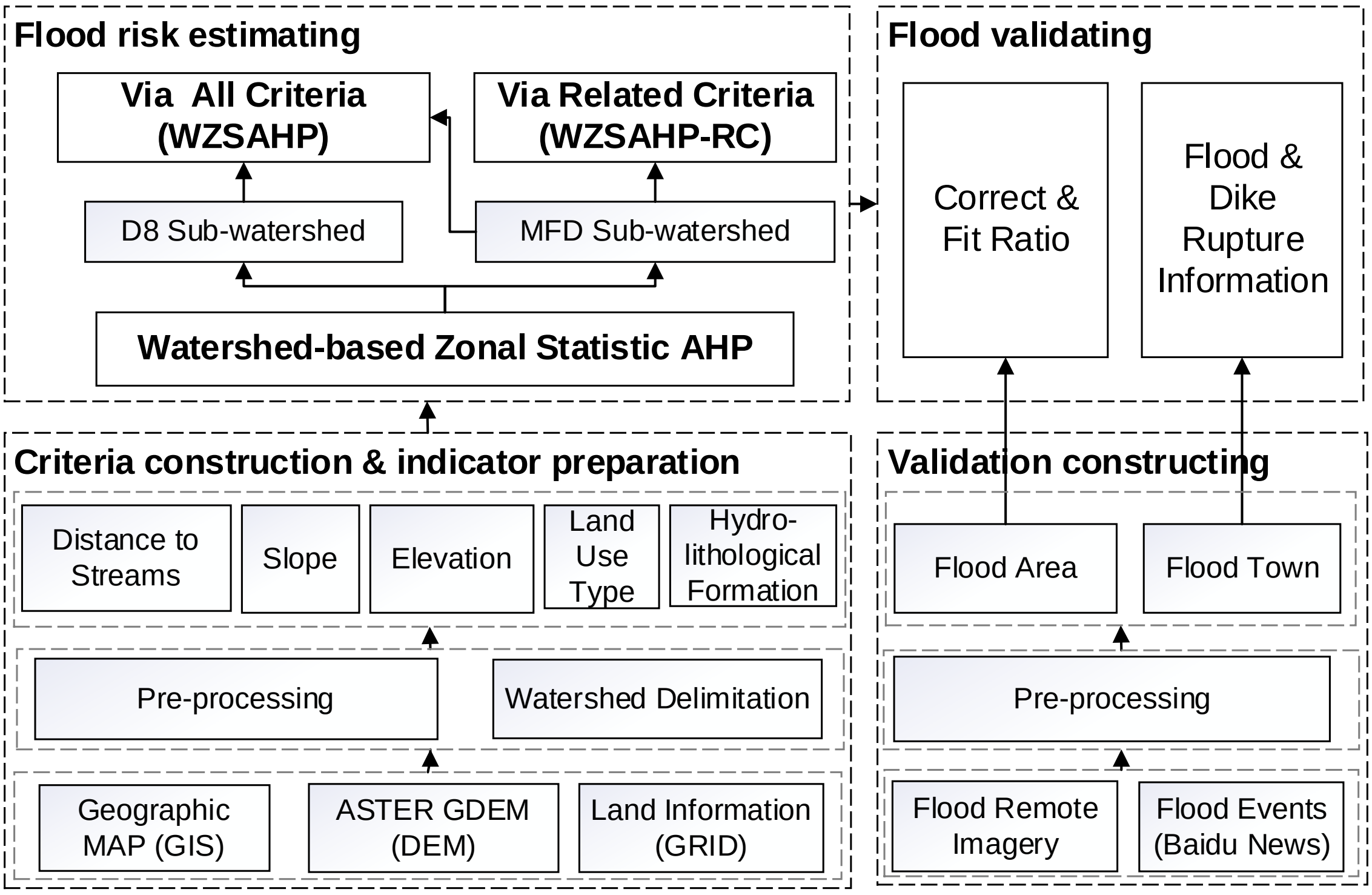

Figure 2. The overall workflow of our study.

The left part of Fig.2 are the flood risk estimation and criteria dataset preparation. Our proposed watershed-based zonal statistic AHP model will be introduced in sub-section 3.1. Based sub-watershed division, we classified WZSAHP as D8-derived and MFD-derived groups, and the MFD-derived group also used to constrain part of related indicators (WZSAHP-RC). The flood risk estimation criteria and indicator preparation will be described in sub-section 3.2, including criteria construction, DEM pre-processing, sub-watershed delaminating, and rating original indicators. The right part of Fig.2 are the flood validating method and the used validating dataset construction. We will introduce the flood risk validating method in sub-section 3.3, including accuracy indicator calculation methods and validation dataset constructing.

### *3.1. Flood risk estimating method*

In flood risk estimation, the AHP model is widely used to integrate a Geographic Information System (GIS) and remote sensing. The structure of traditional AHP (Fig.3(a)) and the intuitive process among tradition AHP (Fig.3(b)) and our proposed WZSAHP (Fig.3(c)) and WZSAHP-RC (Fig.3(d)) are shown as in Fig.3.

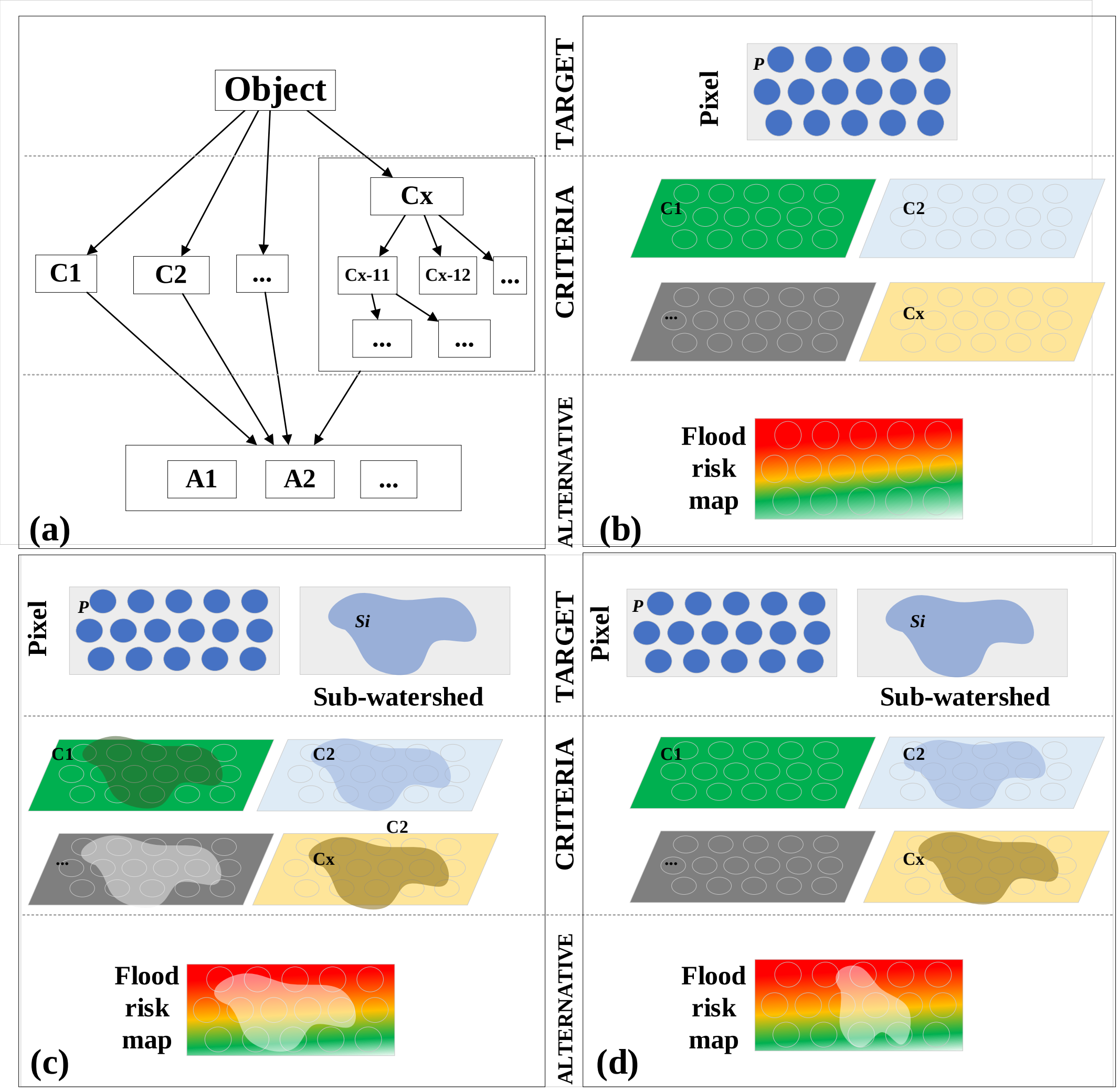

Figure 3. The logical structure of flood risk estimation basing AHP and watershed-based zonal statistical methods: (a) The general structure of AHP, (b)-(d) are Diagram of ordinary AHP, WZSAHP via constraint all criteria, and WZSAHP-RC via constraint related criteria, respectively.

Fig.3 (a) shows the structure diagram of AHP method, its estimation target is $Object$, its criteria is made up as $C = \{c_1, c_2, ..., c_x\}$, and the criterion can be constructed as multiple layers as needed, and its final alternative set $A = \{a_1, a_2, ...\}$ can be determined by the estimating index. Fig.3 (b) shows that common AHP model adopts pixel as a basic unit, the final estimation index can be calculated by accumulating the pairwise cumulative indices and weights. Fig.3 (c) is the diagram of default WZSAHP model, each criterion adopts sub-watershed as basic unit, the original indices use sub-watershed as zonal unit to calculate the maximum, median and mode value of the criteria for each sub-watershed. The final estimation index is calculated as same as Fig. 3(b). Fig.3 (d) is a diagram of WZSAHP-RC model, only part of the runoff indicators is constrained by the sub-watershed. The final estimation index is calculated using the pair wise cumulative criteria and weights. The details of the WZSAHP model are as follows.

AHP is composed of three levels: target, criteria, and alternatives. The target layer refers to the evaluation unit; the criteria (with single or multiple layers) consist of several clusters that reflect different aspects of the target; and the alternative composed of the estimation results set. In the proposed model, the flood risk estimation target, the criteria and the corresponding watershed are defined as follows:

$$P = \begin{bmatrix} p_{11} & p_{12} & \cdots & p_{1n} \\ p_{21} & p_{22} & \cdots & p_{2n} \\ \cdots & \cdots & \cdots & \cdots \\ p_{m1} & p_{m2} & \cdots & p_{mn} \end{bmatrix}, \quad C = \begin{bmatrix} c_1 \\ c_2 \\ \cdots \\ c_x \end{bmatrix}, \quad c_x = \begin{bmatrix} c_{11} & c_{12} & \cdots & c_{1n} \\ c_{21} & c_{22} & \cdots & c_{2n} \\ \cdots & \cdots & \cdots & \cdots \\ c_{m1} & c_{m2} & \cdots & c_{mn} \end{bmatrix}_x \tag{1}$$

where matrix $P$ represents the pixels in study area, with size of $m \times n$, $C$ is the flood risk estimation indicators, and each of the indicator $c_x$ is a raster layer, with size of $m \times n$.

A comparative matrix of criteria and calculated their weights are constructed. According to AHP model, the positive pairwise comparison matrix usage value 1 to 9 indicates the relative importance of two indices, and its largest eigenvalue corresponding eigenvectors can be used as weight vectors to represent the established hierarchic evaluation structure (Saaty, 2004). The hierarchic evaluation structure can be calculated as follows:

$$J = \begin{bmatrix} j_{11} & j_{12} & \cdots & j_{1x} \\ j_{21} & j_{22} & \cdots & j_{2x} \\ \cdots & \cdots & \cdots & \cdots \\ j_{x1} & j_{x2} & \cdots & j_{xx} \end{bmatrix}_{x \times x}, \quad J \bullet X = \lambda_{\max} \bullet X \rightarrow \omega_i = \frac{x_i}{\sum_{j=1}^{x} x_j}, \quad \omega = \begin{bmatrix} \omega_1 \\ \omega_2 \\ \cdots \\ \omega_x \end{bmatrix} \tag{2}$$

where the comparison matrix $J$, with size of $x \times x$, is used to determine the importance order among criteria $C$ (in Eq. (1)). $X$ is the eigenvector corresponding to the largest eigenvalue $\lambda_{\max}$ of $J$, $\omega$ is the weight vector corresponding to the normalization value of eigenvector $X$.

In order to keep the order of relative importance among criteria logically consistent, the consistency ratio as in Eq. (3) was used to calculate judgment matrix (Saaty, 2004). The pair wise comparison matrix comprising can be accepted if its consistency ratio is less than 0.1 (a consistency ratio of 0 indicates that the judgment matrix is completely consistent).

$$CI = \lambda_{\max} - n \big/ n-1, \quad CR = \frac{CI}{RI} \tag{3}$$

where $CR$ is the consistency ratio, $CI$ is the consistency index, $RI$ is a statistic random index, the average $CI$ of randomly generated pair wise of comparison matrix of similar size, $\lambda_{\max}$ is the largest eigenvalue of the comparison matrix, $n$ is the number of indicators used in criteria.

In the proposed WZSAHP-RC model, the final related runoff converging indices value can calculate values for descriptive statistics within sub-watersheds. A sub-watershed is a physical range, it indicates rainwater converging along a section of digital stream flowing out through the same outlet. The range of sub-watershed reflects runoff converge paths and determines the storage of runoff accumulation. The runoff converging related indices, such as slope, and distance from a stream, can constrain sub-watersheds and thus help identify neighborhood risks using following formulas:

$$S = \begin{bmatrix} \cdots & & & \\ & S_k & S_k & \\ & & S_k & \\ & & & \cdots \end{bmatrix}_{m \times n} \tag{4}$$

$$F(S, c_x) = zonalStatistic(S, c_x, Method) \tag{5}$$

where $S$ is the sub-watershed division raster, $F(S, c_x)$ is constraint sub-watershed as statistical zonal unit, to update the corresponding indicator $c_x$, and the size of $F(S, c_x)$ is the also $m \times n$, $zonalStatistic$ is calculated using the descriptive statistics of indicator $c_x$ for each sub-watershed $S$, $Method$ is the statistical method including majority, maximum and median.

The flood risk index can be calculated as in Eq. (6). As the Natural Break slice method can maintain small variance within groups and large variance among index, it is widely used to derive the final flood risk map (Bathrellos et al., 2017; Chakraborty and Mukhopadhyay, 2019; Fernández and Lutz, 2010; Ouma and Tateishi, 2014). Therefore, the final flood risk distribution can be mapped according to flood risk index sliced by the Nature Break method, as {"Very Low", "Low", “Normal”, “High” and “Very High”}.

$$FRI = \omega \bullet C = \sum_{i=1}^{i=m} \omega_i \bullet F(S, c_i) + \sum_{j=m+1}^{j=x} \omega_j \bullet c_j, \ (0 \le m \le x) \tag{6}$$

where $FRI$ is the flood risk index, it can be calculated by the cumulative sum of criteria $C$ and its corresponding weight $\omega$. The criteria can be grouped as the sub-watershed constraint indices $F(S, c_i)$ and the original indices $c_j$.

### *3.2. Criteria construction and indicator dataset preparation*

Flood risk estimation focus on three factors: the flood risk distribution, extreme weather and climate events, and risk exposure in a flood affected area. Among these, the flood risk distribution is the dominant factor, it influenced by its relative geological location and potential quantity of converging rainwater. The hydrological features such as land use type, and porous and impervious features affect local rainfall-runoff ratio, but the runoff will flow through the terrain and drainage network. The terrain features including the slope distribution and the local distance from a stream reflect the path of runoff convergence and potential rainwater accumulation at the watershed scale; these are the main factors influencing the pattern of flood risk distribution.

#### *3.2.1 Constructing flood risk estimation criteria*

In this study, we adopted five indices to construct flood risk estimation criteria, as $C = \{C_1, C_2, C_3, C_4, C_5\}$, where $C_1$= Slope, $C_2$= Elevation, $C_3$= Distance from streams, $C_4$= Hydro-lithological formations, $C_5$= Land use type, reference to (Bathrellos et al., 2017). The definition of comparison matrix also defined reference to the work, as shown in Table 2.

Table 2. The judgment matrix of criteria. $C_1$= Slope, $C_2$= Elevation, $C_3$= Distance from streams, $C_4$= Hydro-lithological formations, $C_5$= Land use type.

| Flood hazard potential | $C_1$ | $C_2$ | $C_3$ | $C_4$ | $C_5$ |
|---|---|---|---|---|---|
| $C_1$ | 1 | 4 | 1/2 | 3 | 1/2 |
| $C_2$ | 1/4 | 1 | 1/3 | 1/2 | 1/4 |
| $C_3$ | 2 | 3 | 1 | 3 | 1 |
| $C_4$ | 1/3 | 2 | 1/3 | 1 | 1/3 |
| $C_5$ | 2 | 4 | 1 | 3 | 1 |

As shown in Table 2, the former three indicators, the slope, elevation, and distance to a stream are indicators affecting the path of rainwater runoff convergence and discharge, they reflect the geological flood risk factors. The hydro-lithological formations and the land use type affect infiltrability, and the roughness of underlying surface and thus the runoff production as expressed in the rainfall-runoff ratio.

In the judgment matrix, the maximum eigenvalue is: $\lambda_{max}$ =5.133. As the criteria number is 5, we obtain the random index value as $RI$ =1.12 from a look up table. Thus, the consistency index of the judgment matrix is: $CI = \frac{\lambda_{\max} - n}{n - 1} = \frac{5.133 - 5}{4} = 0.03325$. Therefore, the final consistency ratio can be calculated as, $CR = \frac{CI}{RI} = \frac{0.03325}{1.12} \approx 0.030$. Since the value of $CR$ is less than 0.1, the judgment matrix used to derive the weight matrix can be accepted. The weight vector can be calculated, as:

$$\mathrm{C} = \begin{bmatrix} C_1 \\ C_2 \\ C_3 \\ C_4 \\ C_5 \end{bmatrix} = \begin{bmatrix} FR_{Slope} \\ FR_{Elevation} \\ FR_{Dis\tan ce_from_streams} \\ FR_{Hydro_litho\log ical_formations} \\ FR_{Land_use_type} \end{bmatrix}, \ A = \begin{bmatrix} 1 & 4 & 1/2 & 3 & 1/2 \\ 1/4 & 1 & 1/3 & 1/2 & 1/4 \\ 2 & 3 & 1 & 3 & 1 \\ 1/3 & 2 & 1/3 & 1 & 1/3 \\ 2 & 4 & 1 & 3 & 1 \end{bmatrix} \quad (7)$$

$$\lambda_{\max} = 5.133, \ \omega = \begin{bmatrix} 0.214 & 0.068 & 0.302 & 0.100 & 0.315 \end{bmatrix}$$

where $C$ is matrix of used flood risk estimation criteria, and the corresponding indicators is: $\left\{ FR_{Slope}, FR_{Elevation}, FR_{Dis\tan ce_from_streams}, FR_{Hydro_litho\log ical_formations}, FR_{Land_use_type} \right\}^{\tau}$,

$FRI$ is the flood risk index, $A$ is the judgment matrix comparing between two indicators, and the weight vector $\omega$ is calculated according to Eq. (2).

As in WZSAHP-RC model, the slope and distance to a stream are runoff converging related indices. The descriptive statistics were calculated, and the flood risk index can be expressed as:

$$FRI = \omega_1 \bullet F(S, C_1) + \omega_2 \bullet C_2 + \omega_3 \bullet F(S, C_3) + \omega_4 \bullet C_4 + \omega_5 \bullet C_5 \tag{8}$$

where $FRI$ is the flood risk index, $\omega$ is weight, $C$ is the criteria as defined according to Eq. (7), $F(S, C_i)$ is the criterion $C_i$ calculated the descriptive statistics values by sub-watershed $S$, according to Eq. (5).

### *3.2.2 Pre-processing watershed data using a DEM and hydrological features*

A DEM was the main data used to derive the watersheds, but need to be enriched with hydrological information before delaminating the sub-watersheds. Kenny and Matthews et al. (2008) found that integrating hydrological streams with a DEM can improve the accuracy in extracting digital drainage. Zhang et al. (2018) pointed out that in the process of delimiting a watershed constraining a DEM by water bodies can reduce the uncertainty created when calculating water flow direction. Thus, hydrological features including streams, rivers, and lakes were used to update the elevation in a corresponding cell of a DEM.

Hydrological features can be recognized by their shapes. Constructed water bodies such as rice paddies and ponds, always have symmetrical regular shapes. Naturally occurring hydrological elements, such as lakes and even constructed elements like reservoirs, always extend along a terrain. Their shapes of these hydrological elements are irregular, and with long perimeters. Other hydrological features such as streams, rivers and ditches, always have long flow path, so their shapes as appear as long narrow rectangles. Therefore, we defined a natural water body index termed Steady Water Index (SWI) as in Eq. (9) to identify hydrological features. For example, river and stream elements, since their shapes are like long narrow rectangles, when the value SWI is 200, the ratio of longer edge and shorter edge is approximately 10, 000. This ratio can represent most rivers and streams.

$$SWI = \frac{Shape_Length}{\sqrt{Shape_Area}} \tag{9}$$

where $SWI$ is the steady water index. The $Shape_Length$ and $Shape_Area$ are the perimeter and area of the water bodies, respectively. Referring to the study area, we suggest the value of SWI should be set as between 6 to 200. At the same time, the water area threshold also used to identify hydrological features. As the minimum area of lakes in the study area is about 70,000 $m^2$, we regarded the water polygons with a spatial coverage larger than 78 cells (as pixel resolution is 30 m, this is about 70,200 $m^2$) as natural water elements.

### *3.2.3 Delaminating sub-watersheds*

We adopted the D8 to delaminate sub-watershed. D8 is a widely used single flow direction (SFD) algorithm. The hydrology tool in ArcGIS 10.3 is based on the D8 algorithm, and was used to segment sub-watersheds in this study. Area threshold of sub-watershed is an important parameter to define watershed schemes. We defined the threshold according to the area of flooded parcels according to Baidu news. The threshold of 200 hectares (ha) was marked according to the reported from network of China Radio (2020), "Wuwei county released flood water to village parcels with area of smaller than 30, 000 mu (~200 hectares) along dikes". The threshold of 667 ha came from Xinhua news (Xinhua News, 2020), "Hefei city flooded nine village parcels with area of larger than 100, 000 mu (~667 hectares) along dikes". There were six area thresholds defined in sub-watershed division as shown in Table. 3:

Table 3. Area threshold used in delimitation watersheds via D8 algorithm.

| Times of | Threshold | (1) | (2) | (3) | (4) | (5) | (6) |
|---|---|---|---|---|---|---|---|
| 667 | unit: ha | 66.7 | | 667.0 | | 3,333.0 | 6,667.0 |
| | unit: mu | ~10,000 | | ~100,000 | | ~500,000 | ~1,000,000 |
| 200 | unit: ha | | 200.0 | | 2,000.0 | | |
| | unit: mu | | ~30,000 | | ~300,000 | | |

As shown in Table.3, four kinds of area thresholds defined according to the area threshold of 667, two kinds of area thresholds defined according to 200. We also used multiple flow direction (MFD) algorithm to divide sub-watersheds, thus the differences influence of flood risk estimation introduced by sub-watershed division derived by SFD and MFD can be compared.

MFD determines the flow direction according to elevation drops between a target cell and adjacent cells. If there are several alternative outflow directions, MFD will choose all of them as the outlets. This may reduce the randomness when setting the flow direction as one of the potential outlets as in flat areas. Referring to our former work (Zhang et al., 2020), watersheds derived by MFD considering terrain connection can maintain higher insistency with flooding points distribution than watersheds derived by SFD as in the D8 algorithm. As in (Zhang et al., 2020), the MFD algorithm calculates flow direction as in Eq. (10), and derives sub-watershed by tracing the flow-in cells.

$$Dir_{flow} += \begin{cases} 2^i, if\ (Z_0 - Z_i) > 0 \\ 0, if\ (Z_0 - Z_i) \leq 0 \end{cases}, \ (0 \leq i \leq 8) \tag{10}$$

where $Dir_{flow}$ is the flow direction of the current cell, it records the potential flow directions in a continuous value between 1–255 by accumulating the potential flow directions; $i$ is the index of the eight adjacent cells, starting from the east, southeast, west, and so on, in a clockwise order; $Z_0$ is the elevation of the center cell; $Z_i$ is the elevation of the adjacent cells.

The MFD algorithm traces the connected flat cells and sets them as the seed of a new sub-watershed. Starting from the range of the seed sub-watershed, all the neighborhood cells flowing into the seed cells will defined as the same sub-watershed. The area will grow, until the area of the traced sub-watershed is larger than the area threshold. The cells traced from the seed sub-watershed will be recorded as a new sub-watershed. Thus, the algorithm can keep the connected flat cells as in a same sub-watershed, while the sub-watershed of the in-flowing neighborhood cells will be determined by the area threshold. The MFD algorithm was programed by C#, and the delimitation is submitted as a supplement materiel as described in Appendix A.

### *3.2.4 Rating flood risk estimation involving indicators*

Flood risk index is calculated by accumulative the value of multiply the indicator with its weight, so the value of original indicators needs to be rated uniformly. The five flood risk indicators are *slope*, *elevation*, *land use type*, *hydro-lithological formation*, and *distance from* *streams*. The value of *slope* and *elevation* is numeric, and their inner distribution is natural. Therefore, we reclassified the *slope* and *elevation* to uniform classes by the natural break method. The value of *land use type* and *hydro-lithological formation* are were ranked according to the corresponding flood risk likelihood level. The values for *distance from streams* were grouped by stream, and the ranking map was labeled according to stream groups. The ranked indicators are listed in Table 4.

Table 4. The classes and rating of factors in flood risk estimation.

| Factors | Classes | Rating | Factors | Classes | Rating |
|---|---|---|---|---|---|
| Slope( °) | 0 | 5 | Land use types | Water | 5 |
| | 0 – 2 | 4 | | Road | 4 |
| | 2 – 6 | 3 | | Building | 3 |
| | 6 – 12 | 2 | | Soil | 2 |
| | 12 - 20 | 1 | | Vegetation | 1 |
| | >20 | 0 | | | |
| Elevation(m) | - 204 - 12 | 5 | Hydro lithological formations | Water | 4 |
| | 12 - 23 | 4 | | Impervious surface | 3 |
| | 23 - 46 | 3 | | Pervious surface | 1 |
| | 46 - 152 | 2 | | | |
| | >152 | 1 | | | |

| Factors | Classes | | | | | Rating |
|---|---|---|---|---|---|---|
| Distance from streams(m) | Rivers, lakes and reservoirs | | | | | 5 |
| | Level 1 | Level 2 | Level 3 | Level 4 | Level 5 | |
| | | | | | 0 – 1,000 | 4 |
| | | | 0 - 500 | 0 – 1,000 | 1,000 – 2,000 | 3 |
| | | 0 - 500 | 500 – 1,000 | 1,000 – 2,000 | 2,000 – 4,000 | 2 |
| | 0 - 500 | 500 – 1,000 | 1,000 – 1,500 | 2,000 – 3,000 | 4,000 – 6,000 | 1 |
| | >500 | >1,000 | >1,500 | >3000 | >6,000 | 0 |

- ***Slope.*** The slope is the main factor influencing rainwater flow path. The slope range is 0 °- 81 °, classified as six classes by Natural Break method, the angle (unit: °) of “0”, “0 - 2”, “2 - 6”, “6 - 12”, “12 - 20”, and “> 20” were labeled as 5, 4, 3, 2, 1 and 0, respectively.

- ***Elevation.*** The elevation influences flood risk distribution. It seems that cells with low elevation have high likelihood suffering flood risk. The elevation range is -204 - 1,490 m, they were classified five type by the Natural Break method, the elevation (unit: m) of "-204 - 12", "12 - 23", "23 - 46", "46 - 152", and ">152" were labeled as 5, 4, 3, 2 and 1, respectively.

- ***Distance from streams.*** Streams are the source of flood risks. The distance from streams reveals the potential risk. In this study, the streams were extracted by D8 algorithm, and the stream levels were labeled by the STRAHLER method. For certain streams, far away cells had lower flood risks than nearby cells. There were six types of distance from streams defined, including water bodies were ranked into 5 classes, and the different distance from stream levels from 1 - 5 were classified as 0-4 according to Table. 4.

- ***Land use types.*** The land use types determine the rainfall-runoff production. We ranked vegetation, soil, building, road and water as 1, 2, 3, 4 and 5, respectively.

- ***Hydro-lithological formations.*** The hydro-lithological formations influence infiltration performed in rainfall-runoff production. Hydro-lithological formations were grouped by water, impervious surface, pervious surface, and they were rated as 4, 3, and 1, respectively.

### *3.3. Flood risk validating method*

In this study, the accuracy of our proposed model was quantitatively evaluated by flooded areas extracted from the GF-3 and Landsat 8 OLI images. We adopted the correct ratio and fit ratio to assess flood estimation accuracy, following Bates and de Roo (Alfieri et al., 2014; Bates and De Roo, 2000). The accuracy can be calculated as follows:

$$Correct(\%) = \frac{FA_{FRI} \cap FA_{Water}}{FA_{Water}} \times 100 \quad (11)$$

$$Fit(\%) = \frac{FA_{FRI} \cap FA_{Water}}{FA_{FRI} \cup FA_{Water}} \times 100 \quad (12)$$

where $Correct(\%)$ is the correct ratio, and $Fit(\%)$ is the fit ratio. $FA_{FRI}$ suggests areas with high likelihood occurring flood. In this study, we construct two validation sets: validation 1: {"Normal", "High", "Very high"} and validation 2: {"High", "Very high"}. $FA_{Water}$ are the flood cells extracted from images. These extracted flooded areas excluded rivers, lakes, ponds, and reservoirs on none-flooding days. The excluded water bodies were identified according to Eq. (9). The correct ratio and fit ratio can distinguish differences improving accuracy in flood or non-flood areas, but a combined estimating indicator is needed to identify the optimal method.

We adopted F1-sorce to distingue the combine accuracy among used methods. F1-score is a common comprehensive indicator in binary segmentation problems. As in binary segmentation problems, more excellent classification usually has higher F1-score. And the value of F1-score is expected to be 0.6~0.8 or higher. Flood risk estimation is a five-types classified question. Therefore, it is acceptable that the F1-score cannot reach the expect range of binary segmentation problems. In this study, we adopted the F1-score values as auxiliary corresponding to correct ratio and fit ratio to identify optimal method. Corresponding to the five-type classification in flood risk estimation, we constructed the comparison matrix according to Table 5:

Table 5. The classes and rating of factors in flood risk estimation.

| $FA_{FRI}$ / $FA_{Water}$ | Validation 1 | | Validation 2 | |
|---|---|---|---|---|
| | Normal, High, Very High | Very Low, Low | High, Very High | Very Low, Low, Normal |
| flooded area | TP | FN | TP | FN |
| dry area | FP | TN | FP | TN |
| normal water | / | / | / | / |

As shown in Table 5, we defined validation 1 and validation 2 according to flood risk level as same as when calculating the correct ratio and fit ratio. The ground truth validation dataset divided as normal water, flooded areas, dry areas. The normal water is the range of hydrological features in non-flooding days. The flood areas are these extracted flooded areas excluded rivers, lakes, ponds, and reservoirs on none-flooding days. The dry areas are the cells exclude water bodies as in the validation dataset. Thus, the F1-score can be calculated according to Eq. (13):

$$P = \frac{TP}{TP + FP}, \quad R = \frac{TP}{TP + FN}, \quad F_1 = \frac{2 * P * R}{P + R} \tag{13}$$

where $F_1$ is the F1-score, $P$ is the precision rate, $R$ is the recall rate, $TP$ is the cells with truthing status while predicted as positive in validation dataset, $FP$ is the cells with false status while predicted as positive in validation dataset, $FN$ is the cells with false status while predicted as negative in validation dataset.

## *4. Results and discussion*

In this session, we will describe the experimental results and discussed our findings. Watersheds deriving by the D8 and MFD algorithms were discussed in subsection 4.1. The area thresholds influence in results for WZSAHP via D8-derived sub-watersheds will be compared in subsection 4.2. Subsequently, differences for WZSAHP-RC via MFD-derived sub-watershed will be provided in subsection 4.3. The correct ratio, fit ratio, and F1-scores among used methods and some conclusion are supplied in subsection 4.4. The subsection 4.5 will discuss the consistency between flood towns with estimation risk.

### *4.1. Comparing the sub-watersheds from D8 and MFD*

Sub-watersheds divided by the D8 and MFD algorithms were mapped by ten discrete colors (Fig.6). As in Fig. 6(a), sub-watersheds delimitated by the MFD algorithm using our programed tool. As in Fig. 6(b)-(g), sub-watersheds produced by the D8 algorithm using hydrological tool in ArcGIS 10.3. The area thresholds of Fig. 6(a) and Fig. 6(b) were 66.7 ha. The area thresholds of Fig. 6(b)-(g) were defined according to Table.3.

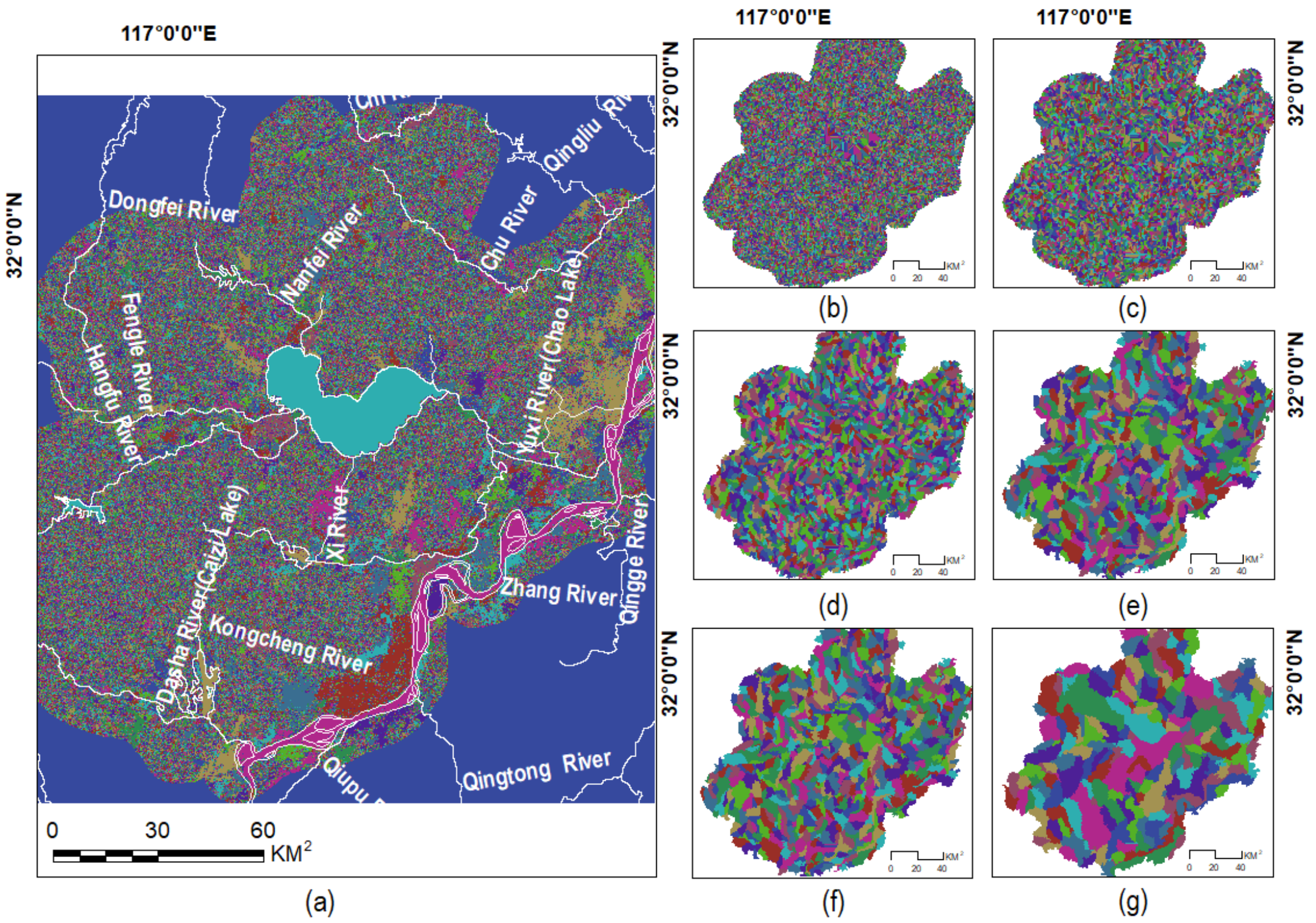

Figure 2. The sub-watersheds derived via MFD (a) and D8 algorithm ((b)-(g)). The area thresholds in (b)-(g) are 66.7 ha, 200.0 ha, 667.0 ha, 2,000.0 ha, 3,333.0 ha and 6,667.0 ha, respectively.

As in Fig.6 (a), the flat area especially the water bodies obviously were kept as in the same sub-watershed. The area did not limit by area threshold. While in Fig. 6(b), all the flat areas and water bodies were treated as common pixels to divide sub-watershed. As in Fig. 6 (b)-(f), the number of sub-watersheds reduced as area threshold increasing. This indicate that the MFD algorithm can keep connecty flat areas as a whole. Thus, if flood taking place in part of flat areas, the sub-watershed derived by the MFD algorithm can recognize connect plat areas as a whole. This might be the reason using MFD-derived sub-watersheds to estimate flood risk may get more accuracy result than using D8-dervied sub-watersheds.

### *4.2. The performance of WZSAHP in D8-derived sub-watersheds*

We estimated the flood risk by AHP (Fig.7(a)) and WZSAHP via D8-derived sub-watersheds (Fig.7(b)-(s)), and also labeled their correct ratio and fit ratio. The correct ratio and fit ratio were calculated by validation 2 as described in sub-section 3.3.

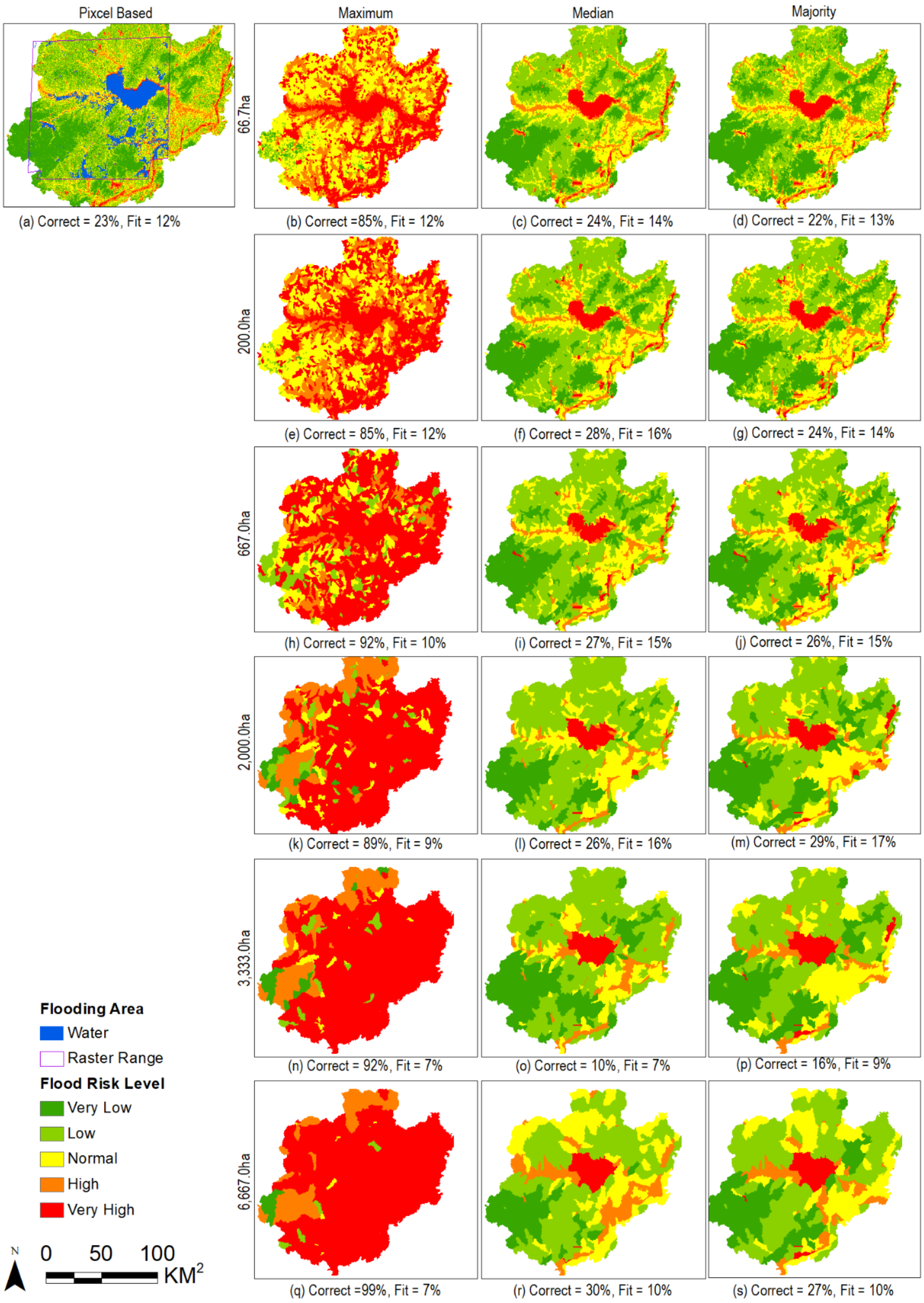

Figure 3. Flood risk levels from pixel-based AHP (a) and AHP based on zonal statistics of sub-watersheds (b)-(s).

As in Fig.7(a), the purple line showed the range of validation remote sensing images, the blue pixels were the flood areas. As in Fig.7(b)-(s), figures in the same row adopted the same area of threshold, figures in the same column used the same statistical method. As in Fig.7 (b), (e), (h), (k), (n), and (q), the flood risk was estimated by maximum statistical method. When using the maximum statistical method, the correct ratio increased obviously (at least increase by 62%, from 23% to 85%), but the fit ratio did not increase, even reduced 5% (from 12% to 7%). In Fig.7 (c), (f), (i), (l), (o), and (r), the flood risk was estimated by median statistical method. As in median statistical method, when area thresholds were 66.7 ha, 200 ha, 667 ha and 2, 000 ha, the correct ratio increased by 1~5%, and the fit ratio increased by 2~4%. However, when the area thresholds were 3, 333 ha and 6, 667 ha, all of the fit ratio reduced. In Fig.7 (d), (g), (j), (m), (p), and (s), the majority statistical method was used to estimate flood risk. As in majority statistical method, the correct ratio and fit ratio increased by 1~6% and 2~5%, respectively, exclude as the area thresholds of 66.7 ha and 3, 333 ha. The fit ratios of area threshold 3, 333 ha and 6, 667 ha reduced when using majority (Fig.7(p)(s)) and median (Fig.7(o)(r)) statistical methods. It indicated that as in WZSAHP, when using median and majority statistical methods, the improvements of correct ratio and fit ratio would be influenced by the area thresholds of D8-derived sub-watershed. And using maximum statistical method to calculate criteria could only improve the correct ratio.

In order to explore the differences between flood risk index distribution derived by AHP ($FRI_{AHP}$) and WZSAHP ($FRI_{WZSAHP}$), we drew the scatter diagrams using flood index point $(FRI_{WZSAHP}, FRI_{AHP})$ corresponding to the flood water cells as in Fig. 8. The points were sampled by a moving $20 \times 20$ grid according to validation range, only the flooded water cells would be chosen. The flood risk index points were drawn according to unique value of $(FRI_{WZSAHP}, FRI_{AHP})$. Each point would be drawn as a transparent circle with variable radius $(alpha = 0.1, radius = 2 * Num_{samepoints})$. As: the alpha value was 0.1, the basic radius was 2, the final variable radius of each unique flood risk point was calculated by the quantity of same unique value multiplied by the basic radius.

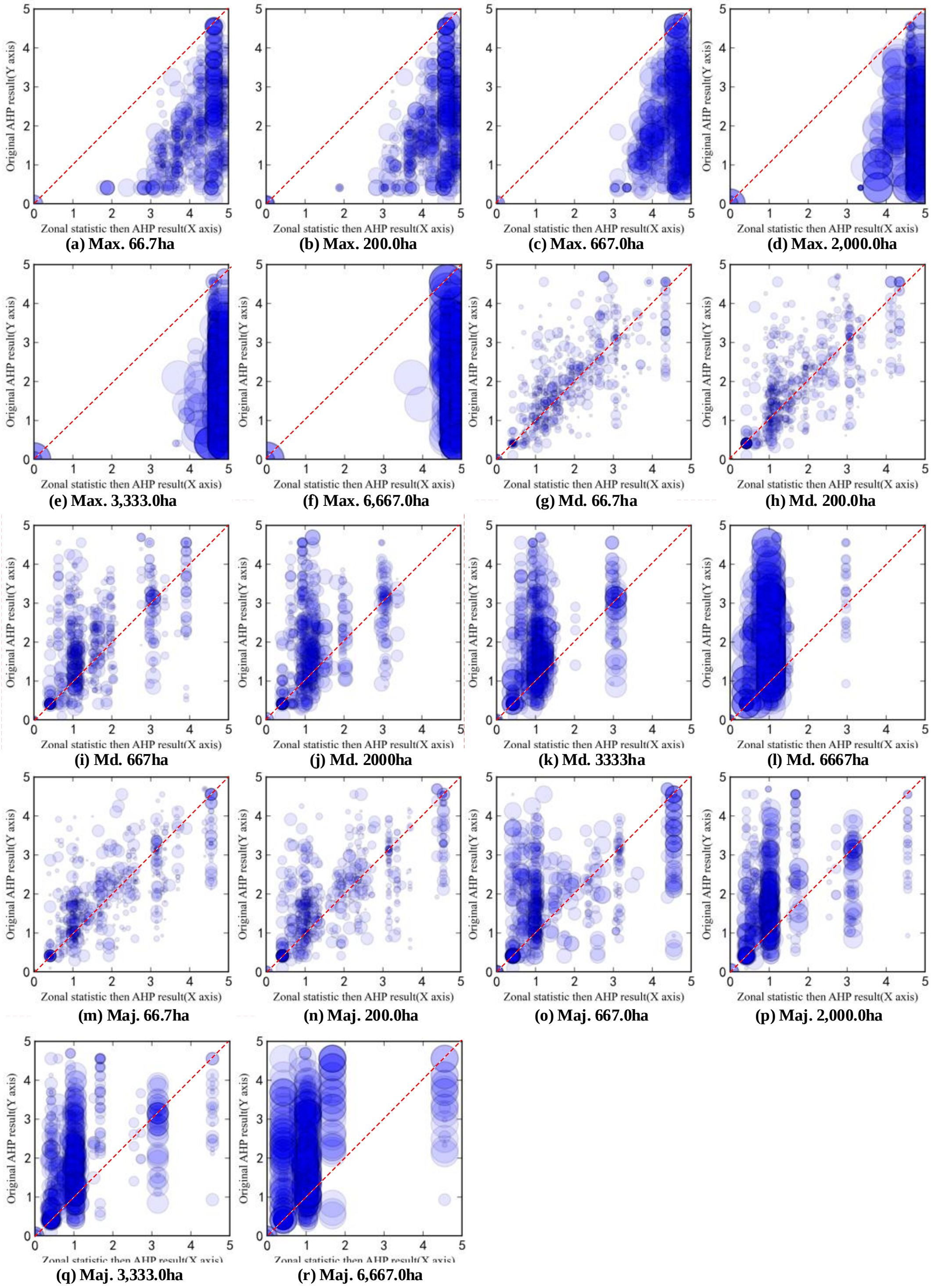

Figure 4. Scatter diagrams of flood risk index derived by WZSAHP via D8 delimitated sub-watershed as X axis and flood risk index derived by AHP as Y axis. Figures (a)-(f) use the "maximum" zonal statistic method, Figures (g)-(l) use the "median" zonal statistic method, and figures (m)-(r) uses the "majority" zonal statistic method.

In Fig. 8, all the sampled flood risk points were true flood water cells. Thus, both the flood risk index derived by AHP and WZSAHP were expected to have a higher value. In Fig.8 (a)-(f), the flood index mapped to X axis ( $FRI_{WZSAHP}$ ) were calculated by WZSAHP model using maximum statistical method, Fig.8 (g)-(l) were calculated by WZSAHP model using median statistical method, and Fig.8 (m)-(r) were calculated by WZSAHP model using majority statistical method. As shown in Fig.8 (a)-(f), when using maximum statistical method, all the points distributed under the 45° diagonal line. This indicated that flood risk index derived by WZSAHP were higher than AHP derived values when using maximum statistical method. As in majority or median statistical method, most points distributed over the 45° diagonal line, especially in Fig.8(k), (l), (o), (r). This means that, when using majority or median statistical methods, the flood risk index derived by WZSAHP lower than AHP. This may lead to flood risk index derived by WZSAHP exist underestimated flood risk likelihood.

### *4.3. The performance of WZSAHP in MFD-derived sub-watersheds*

We compared the flood risk map derived by AHP and WZSAHP via MFD-derived sub-watersheds in Fig. (9). Fig. (9) mapped as same as Fig. (7).

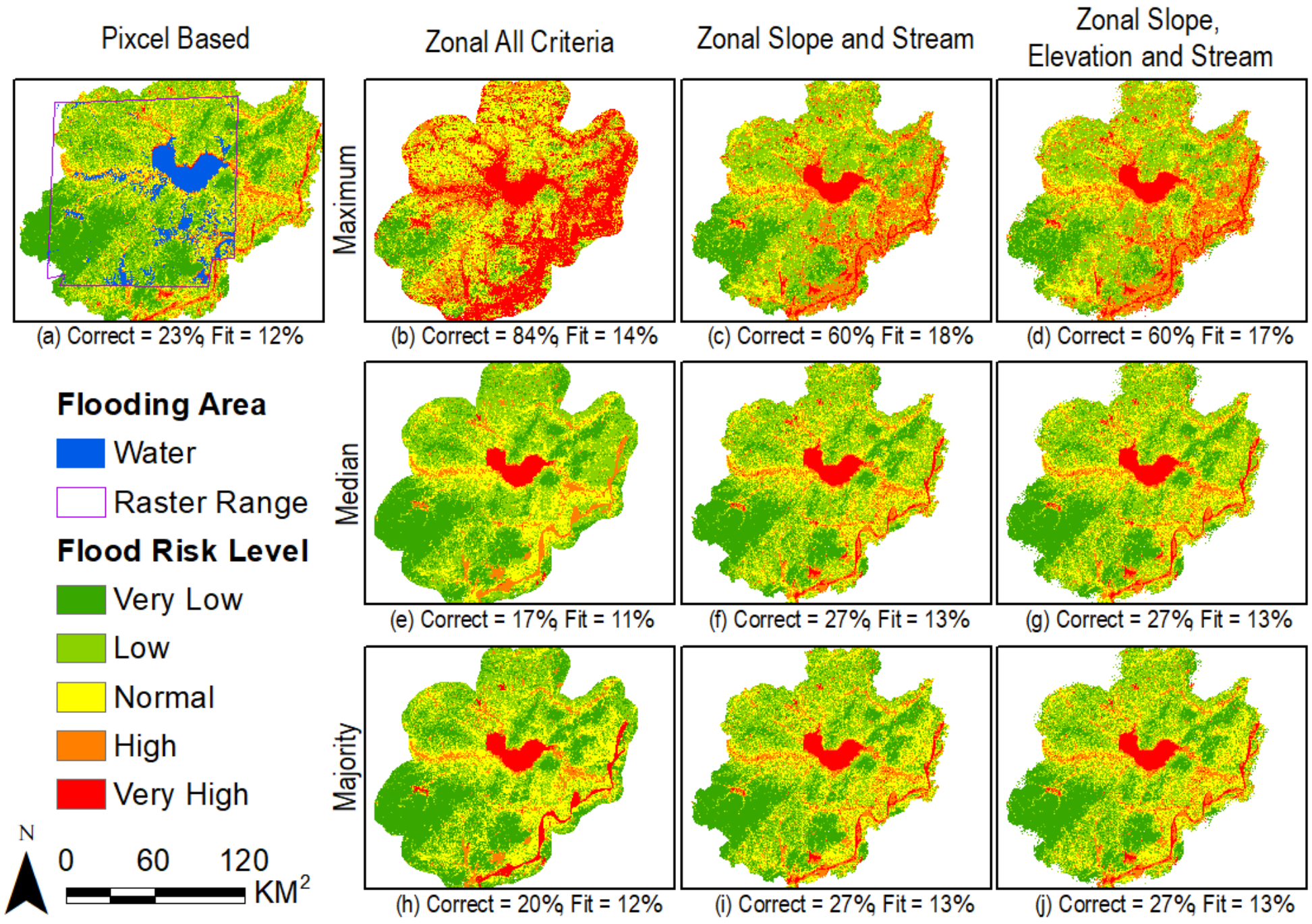

Figure 5. Flood risk levels from pixel-based AHP (a), while (b)-(j) show flood risk levels using different indicators and zonal statistics.

In Fig.9, ground-truthing flood area were colored in blue, flood risk was sliced into five levels. As in Fig.9(b)-(j), figures in the same row adopted the same statistical method, figures in the same column used MFD-derived sub-watersheds to constraint the same kind of related indicators. It was found that, as using MFD-derived sub-watersheds to constraint *slope* & *stream* and *slope* & *elevation* & *stream*, all the correct ratio and fit ratio would be increased. And when using maximum statistical method, the correct ratio and fit ratio obviously increased by 37%~61% and 2%~6%, respectively. When using the maximum statistical method, constrained sub-watershed *slop*e & *stream* could get higher fit ratio than constraint *slope* & *elevation* & *stream,* while both correct ratios got the same value.

We can find that the contributing indicators were the runoff converging related indicators. The *slope* indicator reflects rainwater converging path, the *stream* indicator demonstrates runoff converging area, and the *elevation* indicator shows a pixel belonging to valley, plains, slopes, or plateau, which means the flood likelihood of situ cell. This indicated that using MFD-derived sub-watersheds to constraint runoff converging related indicators can improve flood risk estimation accuracy by using maximum statistical method.

As same as Fig.8, we drew the scatter diagrams to compare the flood risk index value distribution by AHP ( $FRI_{AHP}$ ) and WZSAHP-RC ( $FRI_{WZSAHP-RC}$ ) as shown in Fig. 10. And the scatter points $(FRI_{WZSAHP-RC}, FRI_{AHP})$ mapped flood risk index derived by WZSAHP-RC as X-axis, mapped flood risk index derived by AHP as Y-axis. The sample method to get flood index pair also using a moving grid with size of $20 \times 20$. And the unique index points were also drawn by transparent circles with variable radius $(alpha = 0.1, radius = 2 * Num_{samepoints})$.

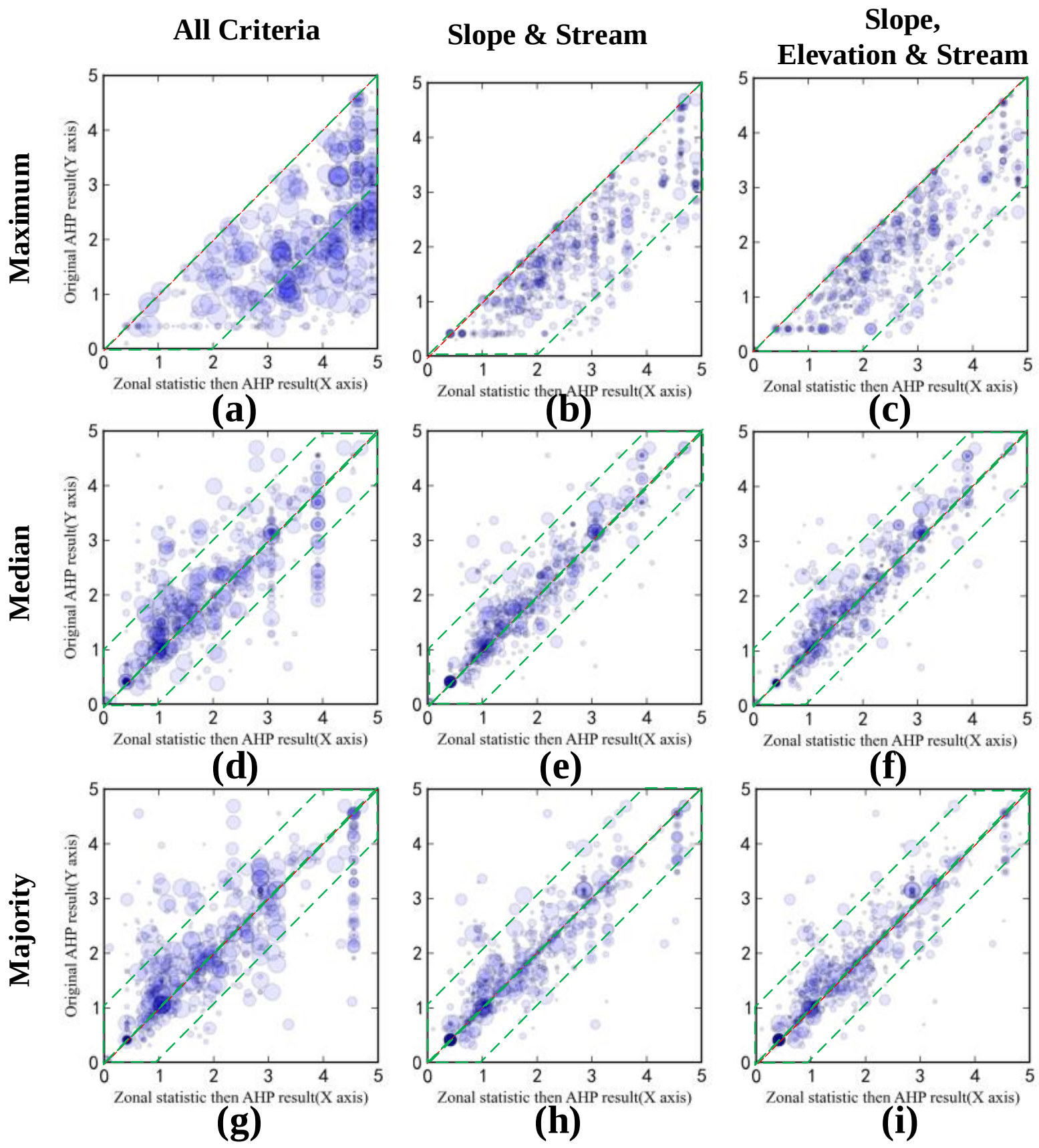

Figure 6. Scatter diagrams of flood risk index derived by WZSAHP via MFD delimitated sub-watershed as X axis and flood risk index derived by AHP as Y axis. The sub-figures in the same row adopt the same zonal statistic method, as (a)-(f) use maximum zonal statistic method, (g)-(l) use median zonal statistic method, and (m)-(r) adopt majority zonal statistic method.

As shown in Fig.10, when using MFD-derived sub-watersheds to constraint *slope* & *stream* (Fig.10(b)(e)(h)) and *slope* & *elevation* & *stream* (Fig.10(c)(f)(i)) indicators, most points mapped in a range of rectangle along diagonal line. As in Fig.10(b)(c), when using maximum statistical method, the points was distributed in the range of rotated rectangle along diagonal line between 0~2. As in Fig.10(e)(f)(h)(i), when using the median and majority statistical methods, the rectangle width was between -1~1 rotated along diagonal line. Meanwhile, as using MFD-derived sub-watersheds to constraint all indicators (Fig.10(a)(d)(g)), the flood risk points mapped more dispersed than using MFD-derived sub-watersheds to constraint related indicators (Fig.10 (c)(f)(i) and (b)(e)(h)). This indicated that the WZSAHP-RC using maximum statistical methods had steady higher estimation value than others. Thus, the WZSAHP-RC might be the optimal method.

### *4.4. Quantitative estimation of the accuracy and fit in AHP and WZSAHP*

The validation 1 and validation 2 described in sub-section 3.3 were used to estimate the accuracy of used models. F1-score was also adopted as auxiliary indicator to assess the comprehensive accuracy. In this study, comparing with AHP, there were six area thresholds of D8-dervied sub-watersheds and MFD-derived sub-watersheds used to constraint indicators by maximum, median, and majority statistical methods, respectively. And there were six kinds of related indicators used to calculate statistical value via deceptively methods constrained by MFD-derived sub-watersheds. The final correct ratio, fit ratio, and F1-Score of AHP, WZSAHP and WZSAHP-RC experiments were listed as in Table 6.

Table 6. The detail correct ratio, fit ratio and F1-score of comparing methods using in this study.

| Adopted Methods | | | Validation 1: {"Normal", "High", "Very High"} | | | Validation 2: {"High", "Very High"} | | |
|---|---|---|---|---|---|---|---|---|
| Statistics | Zonal Method | Base Unit | Cor.1 (%) | Fit1 (%) | F1-Score | Cor.2 (%) | Fit2 (%) | F1-Score |
| None | AHP | Pixel-Based | 67 | 13 | 0.237 | 23 | 12 | 0.216 |
| Max. (MFD & D8) | WZSAHP (MFD) | All Criteria | 93 | 8 | 0.153 | 84 | 14 | 0.249 |
| | WZSAHP-RA (MFD) | Slope | 69 | 14 | 0.251 | 21 | 11 | 0.197 |
| | | Slope & Stream | **83** | **14** | **0.240** | **60** | **18** | **0.299** |
| | | Slope & Elev. | 70 | 14 | 0.241 | 21 | 11 | 0.194 |
| | | Slope, Elev. & Stream | 87 | 13 | 0.231 | 60 | 17 | 0.283 |
| | | Stream | 78 | 14 | 0.253 | 46 | 18 | 0.300 |
| | | Stream & Elev. | 79 | 14 | 0.248 | 47 | 17 | 0.296 |
| | WZSAHP (D8) | 66.7 ha | 99 | 7 | 0.130 | 85 | 12 | 0.212 |
| | | 200 ha | 99 | 7 | 0.130 | 85 | 12 | 0.212 |
| | | 667 ha | 100 | 7 | 0.126 | 92 | 10 | 0.188 |
| | | 2000 ha | 99 | 7 | 0.138 | 89 | 9 | 0.163 |
| | | 3333 ha | 98 | 7 | 0.137 | 92 | 7 | 0.139 |
| | | 6667 ha | 100 | 7 | 0.128 | 99 | 7 | 0.127 |
| Md. (MFD & D8) | WZSAHP (MFD) | All Criteria | 67 | 14 | 0.251 | 17 | 11 | 0.201 |
| | | Slope | 67 | 14 | 0.239 | 25 | 13 | 0.224 |
| | | Slope & Stream | 70 | 13 | 0.233 | 27 | 13 | 0.230 |
| | | Slope & Elev. | 69 | 13 | 0.230 | 26 | 13 | 0.227 |
| | WZSAHP-RA (MFD) | Slope, Elev. & Stream | 69 | 13 | 0.230 | 27 | 13 | 0.229 |
| | | Stream | 66 | 13 | 0.237 | 23 | 12 | 0.218 |
| | | Stream & Elev. | 66 | 13 | 0.236 | 23 | 12 | 0.217 |
| | | 66.7 ha | 68 | 15 | 0.261 | 24 | 14 | 0.240 |
| | | 200 ha | 69 | 15 | 0.266 | 28 | 16 | 0.271 |
| | WZSAHP (D8) | 667 ha | 67 | 15 | 0.266 | 27 | 15 | 0.264 |
| | | 2000 ha | 61 | 15 | 0.255 | 26 | 16 | 0.274 |
| | | 3333 ha | 50 | 12 | 0.215 | 10 | 7 | 0.124 |
| | | 6667 ha | 54 | 10 | 0.174 | 30 | 10 | 0.180 |
| Maj. (MFD & D8) | WZSAHP (MFD) | All Criteria | 73 | 14 | 0.247 | 20 | 12 | 0.209 |
| | | Slope | 61 | 14 | 0.239 | 17 | 10 | 0.190 |
| | | Slope & Stream | 72 | 13 | 0.237 | 27 | 13 | 0.235 |
| | WZSAHP-RA (MFD) | Slope & Elev. | 61 | 13 | 0.233 | 17 | 10 | 0.189 |
| | | Slope, Elev. & Stream | 72 | 13 | 0.235 | 27 | 13 | 0.234 |
| | | Stream | 66 | 13 | 0.234 | 24 | 13 | 0.222 |
| | | Stream & Elev. | 65 | 13 | 0.234 | 24 | 12 | 0.221 |
| | | 66.7 ha | 66 | 14 | 0.252 | 22 | 13 | 0.231 |
| | | 200 ha | 67 | 15 | 0.255 | 24 | 14 | 0.249 |
| | WZSAHP (D8) | 667 ha | 70 | 15 | 0.259 | 26 | 15 | 0.258 |
| | | 2000 ha | 66 | 14 | 0.249 | 29 | 17 | 0.289 |
| | | 3333 ha | 54 | 13 | 0.237 | 16 | 9 | 0.165 |
| | | 6667 ha | 51 | 11 | 0.197 | 27 | 10 | 0.185 |

* Note:

- ✧ Cor. - Correct, ha – hectare, Max. -Maximum, Md. - Median, Maj. - Majority, Elev. - Elevation.
- ✧ The values highlighted by brown denote values higher than the original ratio values.
- ✧ Rows filled by pink suggest that all the six values equal or higher than the original values. And the final proposed WZSAHP-RC model, using maximum statist method via sub-watershed delimitated by MFD to constraint Slope & Stream indicators (WZSAHP-Slope & Stream), is highlight by red and filled by pink.

As shown in Table.6, the correct ratio, fit ration, and F1-score in validation 1 were 67%, 13%, and 0.237, respectively. And the correct ratio, fit ration and F1-score in validation 1 were 23%, 12%, and 0.216, respectively. In this study, we chose those methods promoting the correct ratio, fit ratio, and F1-score in both of the validation datasets as the candidate methods. Comparing with WZSAHP via D8-derived sub-watersheds, when using median statistical method, the correct ratio, fit ratio, and F1-Score were improved as the sub-watershed area threshold were 66.7 ha and 200.0 ha. And when using majority statistical method, the correct ratio, fit ratio, and F1-Score were improved as the sub-watershed area threshold were 667 ha. As in WZSAHP-RC via MFD-derived sub-watersheds, when using maximum statistical method, the correct ratio, fit ratio, and F1-scores were improved, when using the sub-watershed as zonal to calculate statistics of related indicator *Slope* & *Stream*, *Stream*, and *Stream* & *Elevation*. For the six candidate methods, we calculated the relative increasement of correct ratio, fit ratio, and F1-Score and drew the histogram as shown in Fig.11.

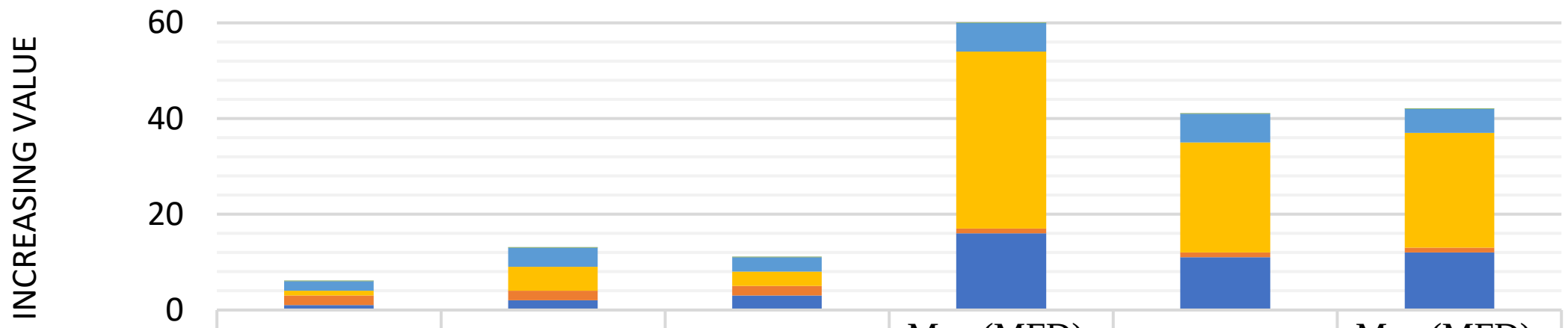

| | Md.(D8)-66.7 ha. | Md.(D8)-200 ha. | Maj.(D8)-667 ha. | Max.(MFD)-Slope & Stream | Max.(MFD)-Stream | Max.(MFD)-Stream & Elev. |
|---|---|---|---|---|---|---|
| F1-Score2 | 0.024 | 0.055 | 0.042 | 0.083 | 0.084 | 0.08 |
| F1-Score1 | 0.024 | 0.029 | 0.022 | 0.003 | 0.016 | 0.011 |
| Fit2 (%) | 2 | 4 | 3 | 6 | 6 | 5 |
| Cor.2 (%) | 1 | 5 | 3 | 37 | 23 | 24 |
| Fit1 (%) | 2 | 2 | 2 | 1 | 1 | 1 |
| Cor.1 (%) | 1 | 2 | 3 | 16 | 11 | 12 |

Figure 7. The increscent correct ratio, fit ratio, and F1-Socre distribution in six candidate methods.

As shown in Fig.11, the increasement of correct ratio and fit ratio in WZSAHP-RC constraint by MFD-derived sub-watersheds were obviously higher than WZSAHP constrained by D8-dervied sub-watersheds. Among the three kinds of WZSAHP-RC model, using MFD-derived sub-watersheds to constraint *Slope* & *Stream* indicators had the highest correctio ratio and fit ratio, and using MFD-derived sub-watersheds to constraint *Stream* had the highest F1-Score. As flood risk estimation can be treated as a five-classification problem, we adopted correct ratio and fit ratio as criteria to choose the proposed method. In this paper, the WZSAHP-RC via maximum statistical method constrained MFD-derived sub-watershed to *Slope* & *Stream* (as WZSAHP-Slope & Stream) was chosen as the proposed method.

### *4.5. Qualitative comparison with the pixel-based AHP and WZSAHP-Slope & Stream model*

We further compared the flood town searched in Baidu News to compare the consistency of flood town between AHP and WZSAHP-Slope & Stream model, as shown in Fig. 12. Fig.12(a)(c) were flood towns overlying the flood risk map derived by AHP, while Fig.12(b)(d) were the flood towns overlying the flood risk map derived by WZSAHP-Slope & Stream model.

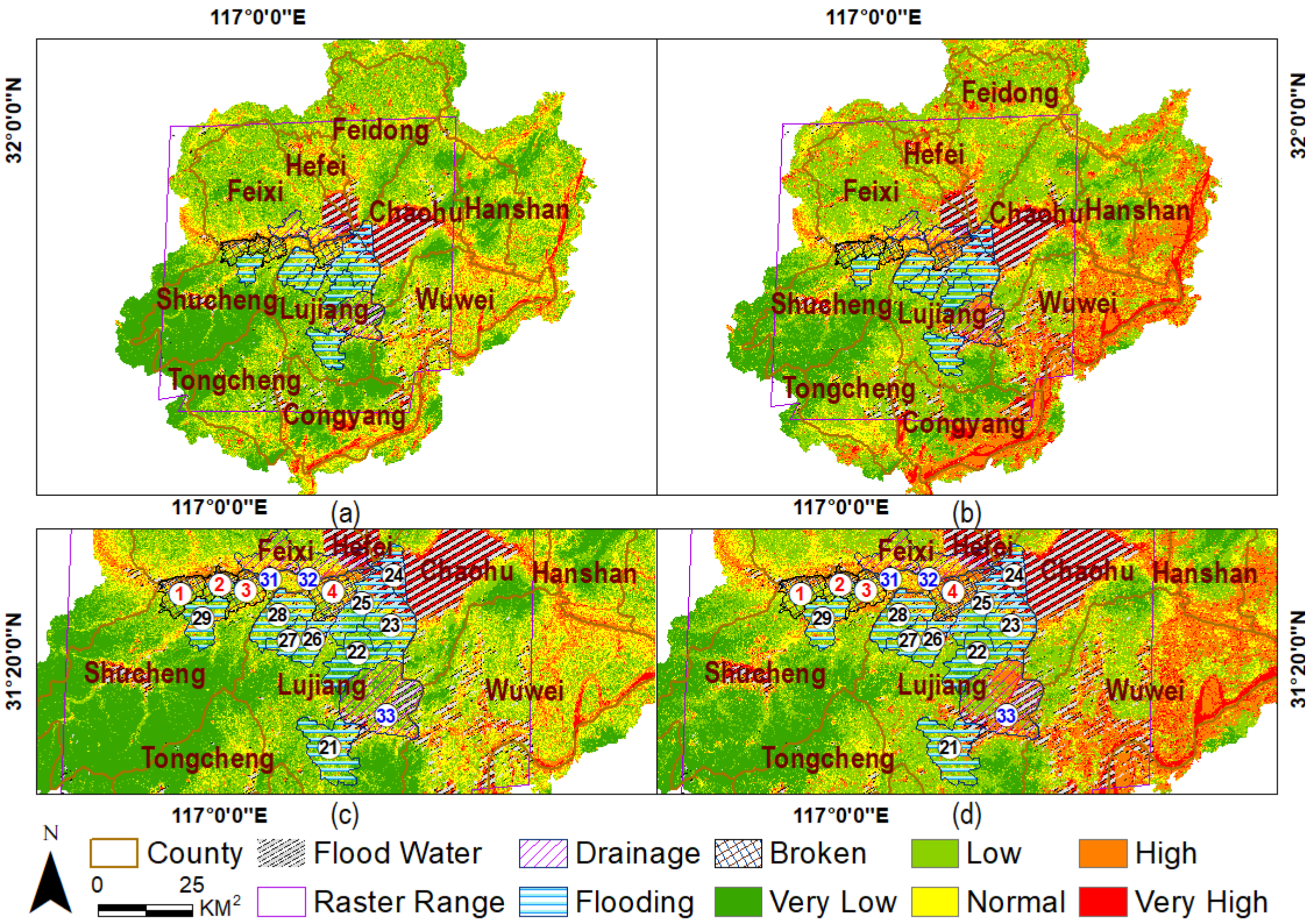

Figure 8. Differences in ground-truthing flooded areas compared with expected flooded areas from the AHP and the WZSAHP constraint by slope and stream. (a) and (c) are from the original pixel-based AHP, while (b) and (d) are from the proposed WZSAHP - Slope & Stream model.

Comparing with the flood risk map, the higher risk areas were obviously distributed along Fengle river, Hangfu river, Zhao river, and the Yangtze River as in our proposed method as in Fig.12(b)(d) than AHP as in Fig.12(a)(c). As shown in Fig.12(c) and Fig.12(d), compared with active drainage areas (labeled as No.31, No.32, and No.33) and dike broken areas (labeled as No.1, No.2, No.3, and No.4), the flood risk levels derived by AHP (Fig.12(c)) were relatively lower than the levels derived by the proposed method (Fig.12(d)). Comparing with the under submerged flooded areas (labeled as No.21~No.29), the flood risk levels distribution derived by AHP and the proposed method were similar. This indicated that the proposed WZSAHP-Slope & Stream method had similar performances in estimating submerged flood areas, while the proposed method had higher consistency with flooded areas caused by dikes broken and active drainage than AHP model.

## *5. Conclusions*

In this study, as traditional pixel-based AHP method failed in capturing the hydrological feature in sub-watershed scale, we proposed a watershed-based zonal statistical AHP constraint to runoff converging related indicators by maximum statistical method (WZSAHP-Slope & Stream) to fill this gap. The proposed method had the highest correct ratio and fit ratio than other methods, including AHP, WZSAHP basing D8-dervied sub-watersheds to constraint criteria, and WZSAHP basing MFD-derived sub-watersheds to constraint other related indicators. We also found that the maximum method could always improve correct ratio. And when using maximum statistical method, all the correct ratio and fit ratio might be improved as constrained MFD-derived sub-watershed to runoff converging related indicators, such as *Slope* & *Stream*, *Stream*, and *Stream* & *Elevation*.

As in the proposed WZSAHP-Slope & Stream model, when using {“Normal”, “High”, “Very high”} as validation set, the correct ratio and fit ratio could increase by 17% (from 67% to 83%) and 1% (from 13% to 14%), respectively. When using {“High”, “Very high”} as validation set, the correct ratio and fit ratio could increase by 37% (from 23% to 60%) and 6% (from 12% to 18%), respectively. And comparing with flood towns filtered by Baidu News, the proposed method had higher consistency with dikes broken and active drainage flooded areas than AHP.

Despite the success of the proposed WZSAHP-Slope & Stream model, we need to acknowledge its limitations. Flooding is a combined event influenced by the disaster-baring environment, disaster-causing factors, and other factors. However, WZSAHP constrained by *Slope* and *Stream* relies solely on indicators, this might lead to its limitation in improving fit ratio. In future research, we plan to combine the carrying capacity of sub-watersheds and the rainwater distribution to further improve the flood risk estimation model.

***Acknowledgments***

This study was supported by the Fundamental Research Funds for the Central Universities (2042021kf0007) and the National Natural Science Foundation of China (41890820, 41771452, 41771454 and 41901340). The authors would like to thank the anonymous reviewers and editors for their comments, which helped us improve this paper significantly. Thanks sincerely to teacher Stephen McClure of Wuhan university for his work in improving this manuscript.

***Appendix A. Supplementary data***

Supplementary data associated with this work can be found in the online version. These data include the rating dataset of flood risk involve of indicators, the watersheds delimitated by D8 (with area threshold of 667 ha, 200.0 ha, 667.0 ha, 2,000.0 ha, 3,333.0 ha, and 6,667 ha) and MFD algorithms, and the water body of natural boundary and during validating flood period.

Table A.1 The details of supplementary data.

| Data name | Data type | Detail | Description |
|---|---|---|---|
| materials. gdb | ARCGIS GDB | R_DEM_5_r_n | The ranked *elevation* indicator. |
| | | R_slope_5_r_n | The ranked *slope* indicator. |
| | | R_IMP | The ranked *Hydro-lithological formations* indicator. |
| | | R_LANDUSE | The ranked *Land use type* indicator. |
| | | R_ED_Stram_water | The ranked *Distance from streams* indicator. |
| | | Watershed_66_7, Watershed_200, Watershed_667, | The watershed derived by D8 algorithm, with area thresholds of 66.7 ha, 200 ha, 667 ha, 2,000 ha, 3,333 ha, 6,667 ha. |
| | | Watershed_2000, Watershed_3333, Watershed_6667 | |
| | | Watershed_MFD | The watershed derived by MFD algorithm. |
| | | Flood_water_body | The water body extracted during flood. |
| | | Mask_water_body | Normal water body extracted in non-flooding days. |
| submit-detailCell Statistics. xlsx | EXCEL | The flood risk accuracy calculation results and the detail flood risk level distribution corresponding to flood or non-flood cells. | |

***Appendix B. The used tiles of impervious surface product***

As shown in 2.2, the 18 tiles (Table B.1) of China's impervious surface product (2m) in Hefei, Luan, Anqing, Wuhu, Maanshan, Chuzhou, Huainan cities were used to prepare the land use and hydrological indicators.

Table B.1. Tiles of China's impervious surface product (2m) used in the study area.

| District | Data tiles | Districts | Data tiles |
|---|---|---|---|
| Anqing | R1C1, R1C2 | Luan | R1C2, R3C2 |
| Chizhou | R1C1, R1C2 | Maanshan | R1C1, R1C2 |
| Chuzhou | R2C1, R2C2 | Tongling | R1C1 |
| Hefei | R1C2, R1C2, R2C1, R2C2 | Wuhu | R1C1, R2C1 |
| Huainan | R2C1 | | |

***Appendix C. The flood town derived from Baidu News***

As shown in section 2.2, we manually cleaned up retrieved raw flood event related news to derive the final validating dataset of flood town (Table C.1) and active breaking dikes for drainage excessive rainwater in Chaohu basin (Table C.2).

Table C.1. The flood- and damage-relevant information from Baidu News in the study area.

| Date | City | Flooded town (village) | Broken location |
|---|---|---|---|
| July 19, 2020 | Shuchengxian, Liuan city | Taoxi town | Fengle River (Longtan River) |
| | | Blinding (Bolin, Jiehe village) | Fenagle (Bolin Reach) |
| | | Qianrenqiao town (Shuxin, Xingfeng, Tonggui, Huangcheng, Wanghe, Sanchahe, Qiandashan etc. villages) | Hangfu, Qiandashan rivers |
| | | Chengguan town (Taiping village) | Sanli, Zhanggongdang, Zhucao Rivers |
| July 22, 2020 | Lujiangxian, Hefei city | Tongda town (Xuejiayu, Guyu, Lianhe, Yongxing, Shifeng, Changfeng villages)<br>Baishan town (Baishan, Daiqiao, Shilian, Jinsheng, Jiulian, Xingang villages) | Shidayu |
| Not clear | | Shengqiao town, Yefushan town, Shitou town, Jinniu town, Guohe town, Nihe town, Baihu town | |

Table C.2. Flooding area distribution caused by active dam breaking.

| Date | City | Flooded land | Mitigation pressure |
|---|---|---|---|
| July 19, 2020 | Quanjiaoxian, Chuzhou city | Dike of Huangcao district 2 and 3 | Chu River |
| July 26, 2020 | Feixixian, Hefei city | Union dike of Jiangkouhe, Yandian Xiang<br>Union dike of Binhu, Sanhe Town<br>Union dike of Shatan, Fengle Town | Chaohu |
| July 27, 2020 | Lujiangxian, Hefei city | Union dike of Peigang, Baihu Town | Chaohu |